\documentclass[12pt]{iopart}
\usepackage{iopams}
\usepackage{setstack}
\usepackage{graphicx}
\usepackage{epsfig}
\usepackage{tensor}
\usepackage[usenames]{color}

\def \be{\begin{equation}}
\def \ee{\end{equation}}
\def \bea{\begin{eqnarray}}
\def \eea{\end{eqnarray}}
\def \ba{\begin{array}}
\def \ea{\end{array}}
\def \nn {\nonumber}

\def\cosh{\mbox{cosh}}

\def \f{\frac}

\def \vp{\varphi}
\def \p{\partial}

\begin{document}

\title[]{Thermodynamics of nonspherical black holes from two-dimensional 
Liouville theory}

\author{Fang-Fang Yuan, Yong-Chang Huang}

\address{Institute of Theoretical Physics, Beijing University of Technology\\
    Beijing 100124, China}
\ead{ffyuan@emails.bjut.edu.cn, ychuang@bjut.edu.cn}
\begin{abstract}
A Liouville formalism was proposed many years ago to account for the black hole entropy. It was recently updated to
connect thermodynamics of general black holes, in particular the Hawking temperature, to two-dimensional Liouville theory.
This relies on the dimensional reduction to two-dimensional black hole metric.
The relevant dilaton gravity model can be rewritten as a Liouville-like theory. We refine
 the method and give general formulas for the corresponding scalar and energy-momentum tensors in Liouville theory.
 This enables us to read off the black hole temperature using a relation which was found about three decades ago.
 Then the range of application is extended to include nonspherical black holes such as neutral and charged black rings, topological black hole and the case coupled to a scalar field.
 As for the entropy, following previous authors, we invoke the Lagrangian approach to central charge by Cadoni and then use the Cardy formula. The general relevant parameters are also given. This approach is more advantageous than the usual Hamiltonian approach which was used by the old Liouville formalism for black hole entropy.
\end{abstract}

\maketitle

\section{Introduction}
A most striking
fact about black hole physics is that one could acquire a lot of information even from the two-dimensional (2D) point of view. Especially the Hawking temperature and entropy of all kinds of black holes have been investigated using many different methods.
Roughly speaking, all
these approaches can be grouped into two main classes: one is based on 2D effective action, and the other relies on holographic 2D conformal field theory.

Back to 1977, Christensen and Fulling \cite{CF77} explored the dynamics of a scalar field in Schwarzschild geometry
by restricting to the $t-r$ plane. They found that the $t-r$ component of the energy-momentum tensor of the resulting effective action gave the Hawking flux and was proportional to the square of the Hawking temperature. In the nineties, other approaches based on effective action have been proposed \cite{MWZ94} - \cite{KV99}.
For new developments, refer to e.g. \cite{BF03, SH08}.

A more systematic method was introduced by Robinson and Wilczek \cite{RW05} in 2005
. They considered a 2D chiral scalar field theory in the near horizon limit of a 4D static
black hole, and found that in order for the anomaly (again, coming from the energy-momentum tensor) to be canceled and the field theory to be unitary, the black hole should behave as a thermal bath with Hawking temperature. This method has been extended to many cases including \cite{IUW06} - \cite{WP1101}.

Ever since the paper of Brown and Henneaux \cite{BH86}, substantial works have been done to study the black holes using conformal field theory. Particularly in 1997, Strominger elucidated the general method and invoked the Cardy formula \cite{CA86} to arrive at the entropy \cite{S97}. Different but related approaches have also been proposed \cite{C98, So98}.
The string theory community interprets this as a holographic correspondence \cite{M97} - \cite{DMW02}.

The past several years witnessed the development of a more powerful method called Kerr/CFT correspondence \cite{S08}.
It showed that even rotating black holes could have 2D (chiral) conformal field theory dual. This also offers a new approach to study other general black holes. 
See \cite{CL11} - \cite{SJY1107} for a partial list of recent developments.

In the framework of 2D dilaton gravity, some earlier works including \cite{RT92, dA92} 
found the relevance
of Liouville-like theory. These are in some sense close to the effective action approach. Later some attempts were made to account for the microscopic degrees of freedom for black hole entropy by studying the emergent Liouville theory at spatial infinity \cite{CHD9506} - \cite {RS0008} (see also a related work \cite{KV0203}).

The near horizon case was investigated by Solodukhin in \cite{So98}, which we will focus on in this paper.
In particular, he considered the dimensional reduction of general gravitational theory to 2D dilaton gravity on the $t-r$ plane, and showed the appearance of Liouville theory (see also \cite{SS9910}) after applying the transformation which was used in \cite{RT92}.
Subsequent developments include \cite{FFGK9901} - \cite{DL0602}. With respect to this old Liouville formalism, several points need to be noted. First, there is already some debate about whether the Liouville theory states represent
the genuine gravitational degrees of freedom. Second, the central charge in this formalism is classical which is from a central term already present in the Poisson brackets. As was found by Carlip \cite{Ca0103} and confirmed by other authors, when Liouville theory is coupled to a dynamical 2D metric, the central charge vanishes. He proposed a method to impose asymptotic fall-off conditions on the metric, either at the infinity or at the horizon.
However, a general relation between the central charge and horizon area need to be borrowed from another approach \cite{C99}. The third point is that the black hole temperature was not obtained through this formalism.

Inspired by these former achievements, the authors in \cite{RY 1} suggested to model the near-horizon limit of general black holes by a 2D Liouville theory. After being transformed into light-cone coordinates, the energy-momentum tensors in the near-horizon regime behave analogously as in \cite{CF77}. So the Hawking temperature can be read off. To obtain the entropy using the Cardy formula, the authors resorted to the results of Cadoni in \cite{Ca}. The latter found explicit relations 
for the central charge and zero mode using the parameters in   the dimensional reduction from $n$D to 2D for black holes.

In \cite{Ca}, Cadoni proposed a Lagrangian approach to central charge and then invoke the
Cardy formula. This is quite different from the usual Hamiltonian approach which was used by the old
Liouville formalism. To evade the difficulty of the Hamiltonian approach to deal with null surfaces, he used a purely Lagrangian formulation of the 2D gravity theory, and derived the central charge from an anomalous term of the transformation of a energy-momentum tensor.
Note that it may be possible to impose horizon constraints in a Lagrangian formalism \cite{Ma0601}, however, this is not the Lagrangian approach that was used by this Liouville formalism.

Similar to the Kerr/CFT correspondence, this renewed formalism can be applied to all kinds of black holes including rotating ones. In \cite{RY 1}, familiar black holes in three and four dimensions have been explored. This method was then extended to study the 4D Kerr-Newman-AdS black hole \cite{RY 2}. Recently, Liouville-type theory was also used to investigate the near-horizon physics of black holes in \cite{Ch10 1,Ch10 2}.

This paper is composed of the following several parts. In the next section, we review the main idea proposed in \cite{RY 1}. In section \ref{S 3}, some formulas are given for general 2D black hole metric which make it straightforward and more easy to apply the method to other black holes. We investigate the black rings and topological black holes, respectively, in 
section \ref{S 4} and section \ref{S 5}.
The Hawking temperature is obtained for each black hole. And the parameters relevant to the entropy as in the Cardy formula and the dimensional reduction are also found. In the last section, we include some comments and suggest further directions.

\section{Black hole/Liouville theory correspondence: A review}
A concrete correspondence between black holes in the near-horizon limit and Liouville theory has been found in \cite{RY 1}. In this section, we will review their formalism to streamline the method.
Particularly, we give the more general formulas which were not explicit in their work. This significantly simplify the use of the approach and make our following discussions more concise.

The main idea of \cite{RY 1} is to model the near-horizon regime of black holes with two-dimensional (2D) Liouville theory. The action reads as follows
\begin{eqnarray}   \label{eq lou}
S_{L}=\frac{1}{96\pi}\int d^2x\sqrt{-g^{(2)}}\left\{-\Phi\square_{(2)}\Phi+2\Phi R^{(2)}\right\}.
\end{eqnarray}
The case with the gauge field is analogous, and was discussed in \cite{RY 2}. Throughout the paper, we use the convention that the Newton's constant $G=1$.

Note that the metric and the Ricci scalar are defined in two dimensions which come from the dimensional reduction of the black hole metric (see the latter part of this section). 
For simplicity, we will omit the superscript "(2)" in the following discussion.
Although the resulting dilaton gravity is generally not conformal, it flows under the renormalization group to
a conformal field theory in the near horizon, which has the form of Liouville theory as in (\ref{eq lou}).

As reviewed in the Introduction, the relevance of Liouville theory to black hole physics has been found in the nineties.
Following \cite{So98}, we consider the dimensional reduction of 4D Einstein-Hilbert action using a spherically symmetric metric of the general form
\begin{equation}
ds^2=\gamma_{ab}dx^adx^b+r^2d\Omega^2.
\end{equation}
Thus an effective 2D theory is obtained
\begin{equation}
S = \int d^2x\sqrt{-\gamma}\left( {1\over 2}(\nabla \Phi )^2+{1\over 4}\Phi^2 R +{1\over 2} \right)~~,
\end{equation}
where $\Phi=r$ can be viewed as a dilaton field, and $R$ is the 2D Ricci scalar. Note that we use conventions  different from that paper.

After applying the transformation
\begin{equation}
\gamma_{ab}=({\phi_h\over \phi })^{1\over 2} e^{{2\over q\Phi_h}\phi} \bar{\gamma}_{ab}~~,~~
\phi={1\over q}{\Phi^2\over \Phi_h}~~,
\end{equation}
the action takes the form of Liouville theory
\begin{equation}
S = \int d^2x\sqrt{-\bar{\gamma}}\left({1\over 2}(\nabla\phi )^2+{1\over 4}q\Phi_h \phi R+U(\phi )\right).
\end{equation}
Here $\Phi_h=r_h$ is the classical value of the field $\Phi$ on the horizon. Note that $\phi_h=q^{-1}\Phi_h$, and $q$ is an arbitrary constant on which the entropy will not depend. As in \cite{So98}, the potential term is also not important for our analysis here. Taking into the freedom of changing $q$, we see that for general black holes it is reasonable to use Liouville theory to study the near horizon physics.

In the following discussions we will review the basic procedure of the formalism in \cite{RY 1}. The energy-momentum tensor for Liouville theory in (\ref{eq lou}) is derived as:
\begin{eqnarray}\label{emt}
\eqalign{
T_{\mu\nu}&=-\frac{2}{\sqrt{-g}}\frac{\delta S_{L}}{\delta g\indices{^\mu^\nu}}\\
&=-\frac{1}{48\pi}\left\{\partial_\mu\Phi\partial_\nu\Phi-2\nabla_\mu\partial_\nu\Phi+g\indices{_\mu_\nu}
\left[2R-\frac12\nabla_\alpha\Phi\nabla^\alpha\Phi\right]\right\}.}
\end{eqnarray}
Note that there are both partial derivatives and covariant derivatives in the above equation which cannot be exchanged.
The equation of motion for the auxiliary scalar $\Phi$ is:
\begin{eqnarray} \label{eom}
\square\Phi=R.
\end{eqnarray}

The 2D metric is obtained following the technique used in the anomaly approach of Hawking radiation proposed by Robinson and Wilczek in \cite{RW05}. 
More detailed examples can be found in \cite{J07} - \cite{PS}.
Generally, the dimensional reduction is implemented by integrating the action of a scalar field in the black hole background. After the angular part is integrated out, we arrive at the 2D metric.

Using this metric, we can then solve the equations (\ref{emt}) and (\ref{eom}), and derive the scalar and the energy-momentum tensor up to integration constants.

The following discussion relies on going to the light-cone coordinates. After that, Unruh Vacuum boundary conditions are adopted \cite{Un76}
\begin{eqnarray}
\label{eq:ubc}
\cases{
T_{++}=0&$r\rightarrow\infty,~l\rightarrow\infty,$\cr
T_{--}=0&$r\rightarrow r_+.$\cr}
\end{eqnarray}
This would determine the remaining integration constants.

Nevertheless, we will concentrate on the near-horizon limit. As shown in the detailed analysis below, one finds that $T^h_{++} = T^h_{--}, T^h_{+-} = T^h_{-+}=0$.
Using the relation (as in \cite{CF77})
\begin{eqnarray} \label{Te}
T^h_{++}=\frac{\pi}{12}{T_H}^2,
\end{eqnarray}
we can read off the black hole temperature.

To discuss the entropy of the black holes, we need to get $g^2_{\mu\nu}$ from $g^D_{\mu\nu}$ through the dimensional reduction,  and use the Cardy formula \cite{CA86}
\begin{eqnarray} \label{Cy}
S=2\pi\sqrt{\frac{c\cdot\Delta_0}{6}}.
\end{eqnarray}
Cadoni explored the case of Schwarzschild black hole \cite{Ca}, and proposed some relations (which were deemed as universal in \cite{RY 1, RY 2}) that could enable us to use the values of relevant parameters to obtain the entropy.



In the paper \cite{Ca}, Cadoni performed the dimensional reduction of gravitational action to obtain a 2D dilaton gravity model as in the work of Solodukhin \cite{So98}.  After a transformation, the action again takes the form of Liouville theory.
For 4D Schwarzschild black hole, the dimensional reduction of 4D Einstein-Hilbert action is done using the metric decomposition as
\bea \label{2D}
ds^2_{(4)} = ds^2_{(2)} + \f{2}{\lambda^2} \phi d\Omega^2,
\eea
where $\lambda$ is the Planck mass.

After Weyl rescaling the 2D metric $g^{2}_{\mu\nu}\to(2\phi)g^{2}_{\mu\nu}$, the 2D dilaton gravity model is given as follows
\bea  \label{sch 2d}
S=\frac{1}{2}\int d^{2}x \sqrt{-g}\left(\phi R +\frac{3}{2} \frac{(\nabla
\phi)^{2}}{\phi}+2\lambda^{2} \phi\right).
\eea
Here $R$ is the Ricci scalar. 

The static solution of this model is
\bea \label{2d ss}
ds^{2}=-\left[(\lambda x)^{2}- \frac{2M}{\lambda} (\lambda x)^{3}\right]dt^{2}+
\left[(\lambda x)^{2}- \frac{2M}{\lambda} (\lambda
x)^{3}\right]^{-1}¥dx^{2}, \\
\quad \phi= \frac{1}{2}(\lambda x)^{-2}.
\eea

A purely Lagrangian method was used there to define the charges associated with the asymptotic symmetry group generators. At last, the central extension and zero mode are given by:
\begin{eqnarray}  \label{Ca r}
c=\frac{48M^2}{\lambda^2},\qquad  \Delta_0=\frac{M^2}{2\lambda^2}.
\end{eqnarray}
So the entropy is as follows
\bea  \label{Ca r2}
S = \frac{4\pi M^2}{{\lambda ^2}}.
\eea

The proposal of the authors in \cite{RY 1} is that the above relations are universal and can be applied to all sorts of black holes. As we see it, so far this is an assumption which can only be justified by the sensible results. For 4D Schwarzschild black hole, $x=\f{1}{r}$, and $\lambda$ is a constant. However, for other black holes, $x$ and  $\lambda$ all depends on the parameters in the 2D metric. These two parameters are crucial for the formalism to account for the black hole entropy.

To be compatible with the 2D metric coming from the anomaly approach as in the former part of the formalism, we will undo the Weyl rescaling $g^{2}_{\mu\nu}\to(2\phi)g^{2}_{\mu\nu}$. So the 2D metric in (\ref{2d ss}) becomes
\bea  \label{2D}
ds^2 =-(1-2Mx)dt^2+(1-2Mx)^{-1}dx^2.
\eea
Compared with Cadoni's convention, it changes back to the usual form for 4D Schwarzschild black hole with $x=\f{1}{r}$.

To sum up, one can use this formalism to derive the corresponding scalar and energy-momentum tensors in Liouville theory. The near-horizon limit of the latter leads to the Hawking temperature. Also we can obtain the entropy and central charge once the parameters in the relations proposed by Cadoni were determined. Although Cadoni's method per se is universal, one may need to be cautious when applying the particular relations in (\ref{Ca r}) and (\ref{Ca r2}) to general black holes. Some comments on this point are included in our Conclusion.

\section{General formulas} \label{S 3}

Since all the 2D metrics in \cite{RY 1} and \cite{RY 2} belong to a class with a uniform ansatz, we find it convenient to derive the general formulas based on it. This will facilitate the analysis of the following sections.

Our starting point is the 2D metric
\begin{eqnarray}   \label{2Dm}
g\indices{_\mu_\nu}=
\left(\begin{array}{cc}-f(r) & 0 \\0 & \frac{1}{f(r)}\end{array}\right).
\end{eqnarray}
We emphasize that one cannot obtain the 2D metric just by discarding other components of the original black hole metric while retaining the $t-r$ components. Instead, we need to integrate the action for a scalar field in the black hole background down to two dimensions. For black rings and topological black holes that we will study in this work, the details can be found in \cite{CH} and \cite{PS}, respectively.

The Ricci scalar can be obtained, which is
\begin{eqnarray}
R= - f''.
\end{eqnarray}

Using the equation (\ref{eom}), we arrive at the expression of the scalar in Liouville theory
\begin{eqnarray} \label{sc1}
\Phi=\int dr \frac{-f'+C_1}{f}+C_2 t+ C_3,
\end{eqnarray}
where $C_1,C_2$, and $C_3$ are all integration constants. One observe that at the horizon
the solution approaches to infinity. It is not clear if this would make the formalism invalid. The physical importance of these constants may need to be investigated further.

To go to the light-cone coordinates, we use the relation $x^{\pm}=2(t\pm r_*)$.
Note that $r_*$ is the tortoise coordinate. The transformation relation is
\begin{eqnarray}
\frac{\partial r}{\partial r_*}=f(r).
\end{eqnarray}
So the light-cone metric is derived as
\begin{eqnarray}
g_{lc}=
\left(\begin{array}{cc} 0 &-\frac{f(r)}{8} \\ -\frac{f(r)}{8} &0\end{array}\right).
\end{eqnarray}

We also have
\bea
dx^{\pm}=2(dt\pm\f{1}{f}dr), \quad \p_{\pm}=\f{1}{2}(\p_t\pm f\p_r).
\eea
From this and equation (\ref{sc1}), the partial derivatives can also be obtained
\bea
\p_+\Phi = \f{1}{2} (-f'+C_1+C_2), \\
\p_-\Phi = \f{1}{2} (f'-C_1+C_2), \\
\p_+^2\Phi = \p_-^2\Phi = -\f{1}{4}ff'', \\
\p_+\p_-\Phi = \f{1}{4}ff''.
\eea

The energy-momentum tensors in the light-cone coordinates are
\begin{eqnarray}
\eqalign{
-48\pi T_{++} = (\partial_+\Phi)^2 - 2 \nabla_+\partial_+\Phi, \\
-48\pi T_{--} = (\partial_-\Phi)^2 - 2 \nabla_-\partial_-\Phi, \\
-48\pi T_{+-} = - 2 \nabla_+\partial_-\Phi + 2 g_{+-}R.}
\end{eqnarray}
Using the expression of the scalar (\ref{sc1}), more usable formulas can be gotten as follows
\begin{eqnarray}
-48\pi T_{++} 
              &=  \frac{1}{4} (C_1+C_2)^2 + \frac{1}{2} ff'' -\frac{1}{4} f'^2,  \label{T 1} \\
-48\pi T_{--} 
              &=  \frac{1}{4} (C_1-C_2)^2 + \frac{1}{2} ff'' -\frac{1}{4} f'^2,  \label{T 2} \\
-48\pi T_{+-} 
              &=  -\frac{1}{4} ff''. \label{T 3}
\end{eqnarray}

When the near-horizon limit is taken, all the integration constants can be ignored, and the ++ component of energy-momentum tensor approaches to
\begin{eqnarray}
-48\pi T^h_{++} = - \frac{1}{4}f'^2 \mid_{r_+}.
\end{eqnarray}
Together with the relation (\ref{Te}), we get the black hole temperature
\begin{equation}
T_H=\frac{f'}{4\pi} \mid_{r_+},
\end{equation}
which is exactly the physical expression.

To obtain this result, one need not to use the formula for the scalar field as in (\ref{sc1}). So that expression only gives a physical value for the corresponding scalar in Liouville theory. However, it may be useful in other developments of this formalism which are unknown to us for now.

The same derivation can also be performed for the following more general 2D metric
\begin{eqnarray}  \label{2Dm 2}
g\indices{_\mu_\nu}=
\left(\begin{array}{cc}-f(r) & 0 \\0 & \frac{1}{g(r)}\end{array}\right).
\end{eqnarray}
After some calculation, the Ricci scalar is obtained as
\begin{eqnarray}
R= - \f{(f'g)'}{2f} - \f{g''}{2} + \f{g'^2}{2g}.
\end{eqnarray}
The equation (\ref{eom}) is difficult to integrate out, so we could not write down the explicit formula for the scalar field in Liouville theory.

In this case, we use the transformation relation
\begin{eqnarray}
\frac{\partial r}{\partial r_*}=\sqrt{f(r)g(r)}.
\end{eqnarray}
And the light-cone metric is found to be
\begin{eqnarray}
g_{lc}=
\left(\begin{array}{cc} 0 &-\frac{f(r)}{8} \\ -\frac{f(r)}{8} &0\end{array}\right).
\end{eqnarray}
Now we have
\bea
dx^{\pm}=2(dt\pm\f{1}{\sqrt{fg}}dr), \quad \p_{\pm}=\f{1}{2}(\p_t\pm \sqrt{fg}\p_r).
\eea

The energy-momentum tensors in the light-cone coordinates can also be derived, which are
\begin{eqnarray}
-48\pi T_{++} &= (\partial_+\Phi)^2 - 2 \partial^2_+\Phi + \sqrt{\f{g}{f}}f'\partial_+\Phi, \\
-48\pi T_{--} &= (\partial_-\Phi)^2 - 2 \partial^2_-\Phi - \sqrt{\f{g}{f}}f'\partial_-\Phi,  \\
-48\pi T_{+-} &= -2\partial_+\partial_-\Phi + 2 g_{+-}R.
\end{eqnarray}
When the near-horizon limit is taken, we expect the $T_{++}$ to approach to
\begin{eqnarray}
-48\pi T^h_{++} = -\f{1}{4} f'g'\mid_{r_+}.
\end{eqnarray}
This is because that combined with the relation (\ref{Te}),
the above equation is consistent with the physical temperature
\begin{eqnarray}
T_H = \f{\sqrt{f'g'}}{4\pi} \mid_{r_+}.
\end{eqnarray}

Most black hole solutions that were found so far have 2D metrics of the form as in (\ref{2Dm}), and exceptional cases are rare. For the black strings, it was noted in \cite{PW08} that the 2D metric, which is not unique, has a more general form as in (\ref{2Dm 2}). However, even in this case, we still have a freedom to choose the metric to be of the form in (\ref{2Dm}).

For general black holes, the dimensional reduction as in equation (\ref{2D}) leads to
the following parameters in Cadoni's relations which determine the entropy through the Cardy formula
\begin{eqnarray}
\lambda^2 = \frac{16\pi M^2}{A}, \\
x = \frac{1-f}{2 M},\\
c = \frac{3A}{\pi},\\
\Delta_0 = \frac{A}{32\pi}.
\end{eqnarray}
Here $A$ is the black hole area. Note that $x$ is a function of $r$ which means that the variable $r$ in $f(r)$ is not set to be the position of horizon. 
The expression of $x$ can be easily read off from our 2D black hole metric ansatz. As for the parameter $\lambda^2$, however, we have to invoke the famous area-entropy law, and then use the relation in (\ref{Ca r2}). One may also note that when we undo the rescaling of 2D black hole metric as did by Cadoni, the $\lambda^2$ disappears in the metric. We are lead to worry about the use of these relations in this formalism. Nevertheless, this point has to be investigated further.

As shown in \cite{RY 1}, the relations given by Cadoni \cite{Ca} lead to central charges whose values are different from the standard research. Using the relation 
 $\f{c_s}{12} = \f{A}{8\pi}$ as in e.g. \cite{C99} ($\f{A}{8\pi}$ is the factor appeared in the conformal algebra), one finds that $c_s = \f{3A}{2\pi}$.  So the standard central charge and zero mode relate to those of Cadoni's as
\begin{eqnarray}
c_s = \f{c}{2},  \\
\Delta_{0,s} = 2 \Delta_0.
\end{eqnarray}

We remark that the detailed investigation of more general dimensional reduction as in \cite{DL0602} is useful. Interestingly, instead of the parameter $\lambda^2$, the authors of that work introduced another parameter $q^2$ (following the original work \cite{So98}, where this parameter was left arbitrary) to obtain the entropy. 

\section{Black rings} \label{S 4}
In this section, we use the refined
formalism 
above to analyze the case of black rings. First of all, we need to obtain the metric in two dimensions as required by the method. This has been done in \cite{CH}. Here we focus on the neutral black ring \cite{ER} which is a vacuum solution of five-dimensional general relativity.

Note that the 2D metric in this section is a solution of the relevant dilaton gravity model which has the form of Liouville theory. However, the detailed dimensional reduction of the 5D gravity as in \cite{DL0602} has not been done. So we will not write down the explicit expression as in (\ref{sch 2d}). We leave this point as one of the open problems for the whole formalism as emphasized in the Conclusion.

The metric of the neutral black ring is given as (cf. \cite{E04})
\bea \nonumber ds^2&=&-\f{F(y)}{F(x)}(dt-CR\f{1+y}{F(y)}d\psi)^2\\
 &\quad&+\f{R^2}{(x-y)^2}F(x)[-\f{G(y)}{F(y)}d\psi^2-\f{dy^2}{G(y)}+\f{dx^2}{G(x)}
+\f{G(x)}{F(x)}d\phi^2], \eea
 with functions
 \be F(\xi)=1+\lambda\xi,
\qquad G(\xi)=(1-\xi^2)(1+\nu\xi), \ee
and the constant \be  \label{C}
C=\sqrt{\lambda(\lambda-\nu)\f{1+\lambda}{1-\lambda}},\qquad
0<\nu\le\lambda<1. \ee
The expressions for the mass and the area of horizon are determined by the parameters $R, \lambda, \nu$.
Explicit formulas are as follows
\begin{eqnarray}
M = \frac{3\pi}{4} R^2 \frac{\lambda}{1-\nu} ,  \\
A = 4 \pi^2 R^3 \frac{\lambda \sqrt{\lambda \nu}}{1-\nu}.
A = 8 \pi^2 R^3 \f{\nu^{3/2} \sqrt{\lambda(1-\lambda^2)}}{(1+\nu)(1-\nu)^2}.
\end{eqnarray}

The coordinates $\psi,\phi$ are two cycles of
the black ring and $x,y$ take values as in
\be -1\le x\le 1,\qquad
-\infty\le y\le -1. \ee
The horizon of the neutral black ring is at $y_H=-\f{1}{\nu}$, and the Hawking temperature is
\be  \label{T nr}
T^{nr}_H 
=\f{1}{4\pi
R}\f{1+\nu}{\sqrt{\lambda\nu}}\sqrt{\f{1-\lambda}{1+\lambda}}. \ee

Near the horizon the action for a scalar field in the black ring background can be integrated down to two dimensions.
This also means that a scalar field theory is reduced to a (1+1) dimensional free field theory.
We expand the scalar field as
\be \label{dre 1}
\vp=\sum_{k,l}\f{1}{2\pi}\vp^{(kl)}(t,x,y)e^{ik\phi}e^{il\psi},\ee
where $k,l$ are integers.
One can further expand function $\vp^{(kl)}(t,x,y)$ using the Legendre polynomial $P_n(x)$ as in
\be \label{dre 2}
\vp^{(kl)}(t,x,y)=\sum_n \vp^{(kln)}(t,y)P_n(x). \ee

With some manipulations as in \cite{CH}, the action can be integrated and written in the following form
\bea  \label{dre 3}
S[\vp]&=&\int dtdy\sum_{k,l,n}\tilde{\vp}^{(kln)}(t,y)
[\f{CR(1+y)}{\sqrt{-F(y)}G(y)}(\p_t+il\f{F(y)}{CR(1+y)})^2 \\
&\quad&-\p_y\f{\sqrt{-F(y)}}{CR(1+y)}G(y)\p_y]\tilde{\vp}^{(kln)}(t,y).
\eea

So the 2D metric we should deal with is
\be
ds^2=-f_{nr}(y)dt{^2}+\f{1}{f_{nr}(y)}dy^2, \ee
where \be \label{f nr}
f_{nr}(y)=\f{\sqrt{-F(y)}}{CR(1+y)}G(y)=\f{(1-y)(1+\nu y) \sqrt{-1-\lambda y}}{C R}, \ee
together with a  $U(1)$ gauge field
\be A_t(y)=-\f{F(y)}{CR(1+y)}. \ee

According to the discussion in the section \ref{S 3}, the corresponding scalar in a Liouville theory can be obtained through equation (\ref{sc1}). The result is
\bea
\Phi_{nr} &=& - \ln f_{nr} + C_1 \int dy \frac{1}{f_{nr}}+C_2 t+ C_3 \nn \\
          & =& - \ln {\frac{(1-y)(1+\nu y) \sqrt{-1-\lambda y}}{CR}} \nn \\
      &&- C_1 \frac{2CR}{1+\nu}
      \left(\frac{1}{\sqrt{1+\lambda}} \arctan \sqrt{\frac{-1-\lambda y}{1+\lambda }}
      + \sqrt{\f{\nu}{\lambda -\nu }} \arctan \sqrt{\frac{\nu (-1-\lambda y)}{\lambda -\nu }}\right) \nn  \\
       &&+ C_2 t+ C_3.
      \eea


Thus we can insert the expression (\ref{f nr}) into equations (\ref{T 1}) - (\ref{T 3}), and find the energy-momentum tensors in the light-cone coordinates
\begin{eqnarray}
-48\pi T^{nr}_{++} =& \frac{1}{16C^2 R^2 (1+\lambda y)} \{ 4(1+\nu)^2 \nn \\
&+ \lambda^2 [3 -2 (1-\nu)y
 + 3 (1+4\nu +\nu^2)y^2 - 6\nu (1-\nu) y^3 -5\nu^2 y^4] \nn \\
& + 4 \lambda [ 1  - \nu + (1+8\nu +\nu^2)y - 3  \nu (1-\nu) y^2 -2 \nu^2 y^3] \}  \nn  \\
& + \f{1}{4}(C_1+C_2)^2,   \\
-48\pi T^{nr}_{+-} =& - \frac{(1-y)(1+\nu y)}{16C^2 R^2(1+\lambda y)} \{8\nu + \lambda^2[1+3(1-\nu)y+15\nu y^2] \nn  \\
&+ 4\lambda(1-\nu+6\nu y) \}. \qquad
\end{eqnarray}
Note that $T^{nr}_{--}$ differs from $T^{nr}_{++}$ only in that the sign in front of $C_2$ is minus. And we have the relation $T^{nr}_{+-} = T^{nr}_{-+}$.
Taking the near-horizon limit $y_H=-\f{1}{\nu}$, the $++$ component of energy-momentum tensor approaches to
\begin{eqnarray}
-48\pi T^{nr,h}_{++} &= - \frac{(\lambda-\nu)(1+\nu)^2}{4C^2R^2\nu} \nn  \\
                &= - \frac{1}{4R^2} \frac{(1+\nu)^2}{\lambda\nu} \frac{1-\lambda}{1+\lambda},
\end{eqnarray}
where equation (\ref{C}) has been used.

Recall the relation (\ref{Te}), so the temperature reads as
\begin{eqnarray}
T^{nr}_H = \f{1}{4\pi
R}\f{1+\nu}{\sqrt{\lambda\nu}}\sqrt{\f{1-\lambda}{1+\lambda}}.
\end{eqnarray}
It is the same as equation (\ref{T nr}) which justifies the formalism.

The parameters in Cadoni's relations can also be obtained as follows
\begin{eqnarray}
\lambda^2_{nr} = \frac{9}{8} \pi R \f{(1+\nu)\lambda^2}{\nu^{3/2} \sqrt{\lambda(1-\lambda^2)}}, \\
x_{nr} = \f{2}{3\pi R^2} \f{1-\nu}{\lambda} [1- \f{(1-y)(1+\nu y) \sqrt{-1-\lambda y}}{C R}],\\
c_{nr} = 24 \pi R^3 \f{\nu^{3/2} \sqrt{\lambda(1-\lambda^2)}}{(1+\nu)(1-\nu)^2},\\
\Delta_0^{nr} = \f{\pi}{4} R^3 \f{\nu^{3/2} \sqrt{\lambda(1-\lambda^2)}}{(1+\nu)(1-\nu)^2}.
\end{eqnarray}
The expressions of $c_{nr}$ and $\Delta_0^{nr}$ are related to that of $\lambda^2_{nr}$ as in Cadoni's relations (\ref{Ca r}). As emphasized in section \ref{S 3}, we make use of the presumed applicability of Cadoni's relations to obtain $\lambda^2_{nr}$ rather than read it directly from the dimensional reduction. This problem need to be resolved in order to solidify this formalism. Some comments about this are also included in the Conclusion.

Using the Cardy formula (\ref{Cy}), the entropy of the neutral black ring is
\begin{eqnarray}
S_{nr} 
= \pi ^2R^3\frac{\lambda \sqrt{\lambda \nu }}{1-\nu }.
\end{eqnarray}


In the case of the single-charged black ring \cite{El03}, the 2D metric differs from the neutral one only by a factor \cite{CH}. The expression is
\begin{eqnarray}
f_{cr}(y)=\f{\sqrt{-F(y)}}{CR(1+y) \cosh^2\alpha}G(y).
\end{eqnarray}
Through the procedure as above, the scalar and the energy-momentum tensors can be found whose forms are very similar to the neutral case. Still only some factors of $\alpha$ appear. Details will not be included here.

In the end, the temperature is found to be the same as the physical expression
\begin{eqnarray}
T^{cr}_H = \f{1}{4\pi R \cosh^2\alpha}\f{1+\nu}{\sqrt{\lambda\nu}}\sqrt{\f{1-\lambda}{1+\lambda}}.
\end{eqnarray}

\section{Topological black holes} \label{S 5}
As black holes in gravitational background of constant negative curvature, 4D topological black hole and the case coupled to a scalar field were analyzed in the framework of anomaly approach for Hawking radiation in \cite{PS}.

The presence of a negative cosmological constant in the action allows black holes with
topology $\mathbb{R}^{2}\times\Sigma$, where $\Sigma$ is a
$(d-2)$-dimensional manifold of constant negative curvature. These
are the so-called topological black holes in \cite{Le94 1} - \cite{
Va}.

We will consider the following simple solution in four dimensions
\begin{equation}
ds^{2}=-f(r)dt^{2}+\frac{1}{f(r)}dr^{2}+r^{2}d\sigma ^{2}, \\
f(r)=\f{r^{2}}{l^2}-1-\frac{2\mu}{r},
\end{equation}
where 
$l$ is the AdS radius, $\mu$ is a
constant proportional to the mass and is bounded from below as
$\mu\geq-\frac{1}{3\sqrt{3}}$. Note that $d\sigma^{2}$ is the line element
of a 2D manifold $\Sigma$, which is locally
isomorphic to the hyperbolic manifold $H^{2}$.

As in \cite{PS}, we will concentrate on the case where $\Sigma$ is a
compact two-dimensional manifold with genus $g=2$.

After dimensional reduction as elaborated there, we get the quite simple 2D metric
\begin{equation}
ds^{2}=-f_{tb}(r)dt^{2}+\frac{1}{f_{tb}(r)}dr^{2}~,
\end{equation}
where
\begin{equation}
f_{tb}(r)=\f{r^{2}}{l^2}-1-\frac{2\mu}{r}.
\end{equation}

Thus we can proceed as in section \ref{S 3}, and the scalar in Liouville theory is
\begin{eqnarray}
  \Phi_{tb}=\int dr \frac{-f'_{tb}+C_1}{f_{tb}}+C_2 t+ C_3   \\
  = -\ln (\f{r^{2}}{l^2}-1-\frac{2\mu}{r}) + C_1 \int dr \f{1}{\f{r^{2}}{l^2}-1-\frac{2\mu}{r}} +C_2 t+ C_3.
\end{eqnarray}
Here we are not able to integrate the second term out in the above equation. 
Compared to this, the case with a scalar field is more easy to tackle. See the latter part of this section.

The following energy-momentum tensors in the light-cone coordinates can also be obtained
\begin{eqnarray}
-48 \pi T^{tb}_{++} =  
\f{\mu(2r+3\mu)}{r^4} - \f{r+6\mu}{l^2 r} +\frac{1}{4} (C_1+C_2)^2,    \\
-48 \pi T^{tb}_{+-} = -\frac{\left( r^3-2l^2\mu \right) \left( r^3-l^2r-2l^2\mu \right) }{2l^4r^4}.
\end{eqnarray}
When the near-horizon limit is taken, $T_{++}$ approaches to
\begin{eqnarray}
-48 \pi T^{tb,h}_{++} = -\left( \f{r_{+}}{l^2}+\frac \mu {r_{+}^2}\right)^2 = -\frac{1}{4}f'^2 \mid_{r_+}.
\end{eqnarray}
So through the equation (\ref{Te}), the black hole temperature can be read off as
\begin{eqnarray}
T^{tb}_H = \frac{f'}{4\pi} = \frac{1}{2\pi}(r_{+}+\frac \mu {r_{+}^2}).
\end{eqnarray}

Using the general formulas in section \ref{S 3}, the parameters in Cadoni's relations is as follows
\begin{eqnarray}
\lambda^2_{tb} 
= \frac{4M^2}{r^2_+}  , \\
x_{tb} 
= \frac{2-\f{r^2}{l^2}+\frac{2\mu}{r}}{2M},\\
c_{tb} 
= 12r^2_+ ,\\
\Delta_0^{tb} 
= \frac{r^2_+}{8}.
\end{eqnarray}
Using the formula (\ref{Cy}), the entropy of the topological black hole is
\begin{eqnarray}
S^{tb} = \pi r^2_+.
\end{eqnarray}

In the above equations, we simply use $r_+$ to denote the position of horizon mainly because it is different in various conditions. More general discussions about topological black holes with different genera, and their positions of horizon can be found in \cite{Va}. Note in passing that the area of horizon for general topological black hole does not have the simple form as $A = 4\pi r^2_+$. (A precise formula is given there as
$A
=4\pi r^2_+(g-1)+\delta(g,1)r^2_+|Im \tau|$, but the detail is irrelevant for us here.)

The case of 4D topological black hole coupled to a scalar field is similar. Now we have
\begin{eqnarray}
f_{tbs}(r) = \frac{r^2}{l^2} - (1+\frac{\mu}{r})^2.
\end{eqnarray}
And the horizon is at the position
\begin{eqnarray}
r_+ = \frac{l}{2} (1 + \sqrt{{1 + \frac{4\mu}{l}}}).
\end{eqnarray}

 Using the general equation (\ref{sc1}), the scalar in Liouville theory can be integrated out to be
\begin{eqnarray}
  \Phi_{tb s}=\int dr \frac{-f'_{tb s}+C_1}{f_{tb s}}+C_2 t+ C_3 \\
      =-\ln[\frac{r^2}{l^2} - (1+\frac{\mu}{r})^2] +\f{C_1 l^2}{\sqrt{4l\mu-l^2}}\arctan \f{2r+l}{\sqrt{4l\mu-l^2}}\nn \\
       -\f{C_1l^2}{\sqrt{4l\mu+l^2}} \arctan h \f{2r-l}{\sqrt{4l\mu+l^2}} +C_2 t+ C_3 .
\end{eqnarray}

The energy-momentum tensors in the light-cone coordinates can also be obtained as follows
\begin{eqnarray}
-48 \pi T^{tbs}_{++} = \frac{2\mu (r+\mu )^3}{r^6}-\frac{r^2+6\mu r+6\mu ^2}{l^2r^2}+\frac{1}{4} (C_1+C_2)^2,  \\
-48 \pi T^{tbs}_{+-} = - \frac{(r^4 - 2 \mu l^2 r - 3 \mu^2 l^2)[r^4 - l^2 (r + \mu)^2]}{2 l^4 r^6} .
\end{eqnarray}
When we take the near-horizon limit, $T_{++}$ is now
\begin{eqnarray}
-48 \pi T^{tbs,h}_{++}  = -\frac{1}{4} {f_{tbs}'}^2 \mid_{r_+},
\end{eqnarray}
where $f'_{tbs} = \f{2r}{l^2} + \f{2\mu(r+\mu)}{r^3} $.
So through equation (\ref{Te}), the black hole temperature can be seen as
\begin{eqnarray}
T^{tbs}_H = \frac{f'_{tbs}}{4\pi} \mid_{r_+}.
\end{eqnarray}
The parameters defined by Cadoni \cite{Ca} in the dimensional reduction from 4D to 2D is also similar, so we omit the explicit expressions here.

We conclude this section with a comment. Interestingly, in 2000, Krasnov \cite{Kr0005} investigated the holography for asymptotically Anti de Sitter (AdS) spaces with arbitrary genus compact Riemann surface as conformal boundary. Such spaces can be constructed from Euclidean AdS3 spaces by discrete identifications, which is like topological black holes we considered above. After calculating the semi-classical gravitational action for each space, the action of Liouville theory was found to be reproduced. The author interpreted the spaces as ¡°analytic continuations¡± of  Lorentzian signature black holes. One may wonder whether there is some connection between that work and this formalism.

\section{Conclusion} \label{S 6}
In this work, we employed the new Liouville formalism proposed in \cite{RY 1} to investigate the thermodynamics of black rings and 4D topological black holes. This provided a new approach to obtain the Hawking temperature and entropy for the nonspherical black holes we studied in this paper. Explicit expressions of the scalar field and the energy-momentum tensors in relevant Liouville theory were also given for general 2D metric. This greatly facilitated our discussions and would be useful for the formalism per se.

The starting point of this method is the corresponding 2D metric for a specific black hole. According to the anomaly approach of Hawking radiation proposed in \cite{RW05}, one can read off the 2D metric by integrating the action for a scalar field in the black hole background down to two dimensions. After dimensional reduction of the gravitational action for the black hole to two dimensions, one obtains a 2D dilaton gravity model which has the form of Liouville theory. Then the energy-momentum tensors can be explicitly written down. When the near-horizon limit is taken, we could invoke the important relation between the energy-momentum tensor and the Hawking temperature first discovered in \cite{CF77} to read off the black hole temperature. The entropy can also be obtained with the help of Cadoni's relations in \cite{Ca}.

One may consider the following further directions. First, the extension to dipole black ring \cite{CH} and black strings \cite{PW08} seems feasible. Nevertheless, for the former case, the integration of the expression for the scalar is difficult. As for the black strings, since the explicit formula for the position of horizon is missing, it is not straightforward to take the near-horizon limit of the energy-momentum tensor. Even so, other more complex black ring solutions are still worthy of further investigation.

Second, black holes in (gauged) supergravity including higher dimensional ones could be investigated using this method (see e.g. \cite{CCLP} and \cite{WP1101}). Note that the 4D Kerr-Newman-AdS black hole has been studied in \cite{RY 2}. 
We perform the dimensional reduction for the 4D Cvetic-Youm black hole, and find the 2D metric with $f=\f{\Delta}{2m[c_{1234} r - s_{1234} (r-2m)]}$. The procedure is the same as in e.g. \cite{CH} (see also the equations (\ref{dre 1}) - (\ref{dre 3})), and we omit the detail here.
The following discussion is
analogous to that in the main text. Since this
is not compatible with
the topic of this paper, we did not report it here. However, other black holes in supergravity are perhaps more difficult to deal with.

Some questions about the formalism itself also deserve to explore. As already hinted in our discussions, at least two problems may be encountered in the applications of this method: first, the expression for the scalar field in Liouville theory is sometimes not easy to integrate out; second, the near-horizon limit may be difficult to take for the energy-momentum tensor. This is sometimes intertwined with the fact that the position of the horizon is not explicitly given.

Some comments on the use of Cadoni's relations \cite{Ca} in this formalism are 
also indispensable here. Firstly, as noted in \cite{RY 1} (see also the end of section \ref{S 3}), the relations proposed by Cadoni lead to central charges whose values are different from the usual research. Although in some sense this is harmless, further exploration following his work is desirable.

Secondly, the strong resemblance between the 2D metrics of static solutions and that of nonstatic solutions may not be a sufficient reason for us to trust Cadoni's relations for general cases, although the sensible results of these works can be seen as a good sign. The authors of \cite{RY 1, RY 2} argued/assumed the applicability of these relations to general black holes, and found reasonable solutions for many kinds of black holes including rotating ones. Note that Cadoni only considered the static solutions of the 2D model coming from dimensional reduction of the 4D Einstein-Hilbert action. One may invoke the analyses of more general dimensional reduction in \cite{DL0602} (cf. \cite{So98, SS9910}) to argue the suitability to apply them to general static solutions. However, to solidify the application 
to rotating black holes and nonspherical ones, one may have to directly study the particular 2D model corresponding to 2D black hole metric coming from the anomaly approach of Hawking radiation \cite{RW05}, and then perform the Lagrangian method of \cite{Ca} to find the central charge and the relations. Another possibility is that there is some convincing argument unknown to us.

Thirdly, just as Carlip's improved version for the old Liouville formalism \cite{Ca0103} has to use the relation $c=\f{3}{2\pi}$ which is from another independent method, we cannot obtain the general formula of the parameter $\lambda^2$ from the new formalism itself (see the end of section \ref{S 3}). This point would also be resolved if we employ the dimensional reduction procedure more directly as in \cite{DL0602}. Apparently, the connection to the old Liouville formalism should also be investigated further.

Compared to the usual approach based on conformal field theory, although this formalism is not very complicated, it has the advantage that explicit action, energy-momentum tensors, and also the scalar field could be given. One may expect that all the information here could facilitate us to extend the method to explore other aspects of black holes. This would be more valuable than the investigations in our paper.

\ack
We thank Sande Lemos, Bibhas Majhi, and Hossein Yavartanoo for kindly correspondence.
This work was supported by National Natural Science Foundation of China (No. 10875009) and by Beijing
Natural Science Foundation (No. 1072005).



\end{document}